\documentclass[journal]{IEEEtran}

\usepackage{mathtools}
\usepackage{graphicx}
\usepackage{array}
\usepackage{url}
\usepackage{titlesec}
\usepackage{float}
\usepackage{enumitem}
\usepackage{cite}
\usepackage{multirow}
\usepackage{todonotes}
\usepackage{bm}
\usepackage{blkarray,bigstrut}
\usepackage{rotating}
\usepackage{amssymb}
\usepackage{tabularx}

\usepackage{textcomp}
\usepackage{multirow}
\usepackage{xcolor}

\usepackage{nomencl}
\makenomenclature

\titlespacing{\section}{0pt}{5pt}{5pt}
\titlespacing{\subsection}{0pt}{5pt}{1pt}

\IEEEoverridecommandlockouts

\setlist{nolistsep}

\begin{document}


\title{Unit Commitment on the Cloud}

\author{Mushfiqur~R.~Sarker,~\IEEEmembership{Member,~IEEE}, and ~Jianhui~Wang,~\IEEEmembership{Senior Member,~IEEE} 


\vspace{-10pt}

}

\maketitle

\begin{abstract}

The advent of High Performance Computing (HPC) has provided the computational capacity required for power system operators (SO) to obtain solutions in the least time to highly-complex applications, i.e., Unit Commitment (UC). The UC problem, which attempts to schedule the least-cost combination of generating units to meet the load, is increasing in complexity and problem size due to deployments of renewable resources and smart grid technologies. The current approach to solving the UC problem consists of in-house HPC infrastructures, which experience issues at scale, and demands high maintenance and capital expenditures. On the other hand, cloud computing is an ideal substitute due to its powerful computational capacity, rapid scalability, and high cost-effectiveness. In this work, the benefits and challenges of outsourcing the UC application to the cloud are explored. A quantitative analysis of the computational performance gain is explored for a large-scale UC problem solved on the cloud and compared to traditional in-house HPC infrastructure. The results show substantial reduction in solve time when outsourced to the cloud.
\end{abstract}


\IEEEpeerreviewmaketitle

\section{Introduction}

The recent paradigm shift in the power system industry is to take advantage of High Performance Computing (HPC) infrastructures to solve operation-based applications, such as Unit Commitment (UC) \cite{ISONERobustUCHPC,papavasiliou2013comparative,Papavasiliou2015}. UC is a highly-complex mixed integer linear program (MILP) which determines the commitment status of participating supply providers (e.g., conventional units, renewable resources, and even demand-side resources) to meet the net demand at each operating bus in the system, while adhering to supply provider and transmission constraints \cite{PandzicUC,Papavasiliou2015}. The need for HPC arises from the increasing amount of renewable resources and smart grid technologies, which has enlarged the computational complexity and problem size of the UC application. Recent developments of solving UC with HPC infrastructure in the open literature (e.g., in \cite{papavasiliou2013comparative,Papavasiliou2015}) and also  in real-life operations (e.g., by ISO-NE \cite{ISONERobustUCHPC}) discuss two major benefits: 1) reduced computation time, and 2) inclusion of mechanisms to consider uncertainty of renewable resources. Traditionally, however, HPC infrastructure is hosted by SOs in local computing environments (e.g., by ISO New England \cite{ISONERobustUCHPC}), where the flexibility to enhance computational capacity becomes a bottleneck, because marginal performance increases require high capital expenditures and maintenance.

On the other hand, cloud computing is emerging as a new paradigm shifting technology to solve complex power grid applications. The outsourcing of UC to the cloud enables tapping into its powerful computational capacity, rapid scalability, and high cost effectiveness. Pioneering system operators (SO) have explored the implementation of power system planning applications on the cloud, such as by ISO New England \cite{Litvinov2016,NewtonGroup2014}. However, limited works exist on analyzing the performance gain of the UC problem on the cloud.

The objective of this work is to introduce the benefits and challenges of outsourcing the UC application to the cloud. The UC is solved using a large-scale power system to analyze the computational performance under different categories of cloud-based instances with comparisons to a traditional in-house HPC infrastructure.

\section{Benefits and Challenges with UC on the Cloud}

The powerful computational capacity, rapid scalability, and high cost effectiveness are the three major benefits to cloud computing that SOs can exploit. Specifically, for example, Amazon's cloud product, EC2, provides up to 128 virtual CPUs and 1952 GB of RAM per cloud instance, which can be provisioned in a short period of time. The ability to outsource simulations to the cloud decreases the requirement for in-house HPC infrastructures, thus providing potential savings in both cost and computational overhead. 

Although there are major benefits of the cloud, there are significant challenges related to the cybersecurity aspects SOs must consider. Three overarching cybersecurity challenges must be addressed for operation-based applications: (1) infrastructure security, (2) data confidentiality, and (3) time criticality. Within (1) infrastructure security, the local and cloud infrastructures, and the communication between them must be secured from potential insider and outsider cyberattackers with intentions of passive (i.e., eavesdrop on the data streams) or active (i.e., maliciously perform false data injections) manipulations. Furthermore, (2) data confidentiality is crucial for operation-based applications, such as UC. The UC problem includes generation- and network-specific data, which must remain confidential and thus mechanisms must be established to secure data from cyberattacks. Lastly, the secure outsourcing of UC on the cloud, must conform to the (3) time criticality requirements set forth by SOs. Their will be inherent tradeoff between enacting enhanced cybersecurity measures and the solve time for UC on the cloud. The challenge remains in finding solution mechanisms that ensure high security and low computational overhead. 


\section{Unit Commitment Formulation}

The compact matrix formulation of the SCUC problem is shown in \eqref{scuc_1}-\eqref{scuc_6}. The objective function \eqref{scuc_1} minimizes the sum of the commitment costs $\textbf{c}^T$\textbf{z} (i.e., start-up, no-load, and shut-down costs) and dispatch costs $\textbf{b}^T\textbf{y}$ over the operating horizon. The binary variable $\textbf{z} \in \{0,1\}$ is a vector of commitment related decisions, including the ON/OFF and start-up/shut-down status of each generation unit within each time interval. The continuous variable $\textbf{y}$ is a vector of dispatch related decisions, including the generation output. Equations \eqref{scuc_2}-\eqref{scuc_6} contain commitment and dispatch related constraints. For a detailed formulation, the interested reader is encouraged to refer to \cite{PandzicUC}.
\begin{eqnarray}
min_{z,y} &\textbf{c}^T\textbf{z}+\textbf{b}^T\textbf{y} \label{scuc_1} \\
s.t. &\textbf{Fz} \leq \textbf{f}, \label{scuc_2} \\
&\textbf{Hy} \leq \textbf{h}, \label{scuc_3} \\
&\textbf{Az}+\textbf{By} \leq \textbf{g}, \label{scuc_4} \\
&\textbf{I}_{u} \textbf{y} = \textbf{d}, \label{scuc_5} \\
&\textbf{z} \in \{0,1\} \label{scuc_6}
\end{eqnarray}

\section{Simulation Results}

\begin{table}[tb]
\renewcommand{\arraystretch}{1.2}
\caption{Computing infrastructure characteristics}
\centering
\begin{tabular}{l| c c c c}
\hline
    &  CPU &  RAM &  SSD &  Intel Processor \\
\hline \hline
1) ANLBlues & 16 & 64 & \checkmark &   Xeon Nehalem  \\
2) c4.2xlarge & 8 & 16 & \checkmark &  Xeon E5-2666v3   \\
3) c4.4xlarge & 16 & 30 & \checkmark &  Xeon E5-2666v3   \\
4) c4.8xlarge & 36 & 60 & \checkmark &  Xeon E5-2666v3     \\
5) m4.16xlarge & 64 & 256 & \checkmark &  Xeon E5-2686v4 \\
\hline \hline
\end{tabular} 
\label{tab:compdetails}
\end{table}

To analyze SCUC performance, the Illinois, USA power system was used, which lies within the Midcontent Independent System Operator's (MISO) region. This system consists of 210 generators, 1908 buses, and 2522 transmission lines. The SCUC model consists of four piecewise linear cost segments to preserve linearity. The SCUC leads to a highly-complex MILP problem with 237,817 variables, of which 55,440 are binary. The model is formulated based on \cite{PandzicUC} using GAMS 24.0.1 \cite{GAMS} and solved using IBM's CPLEX \cite{cplex} solver. The optimality gap was set to 0.5\%.

For computational comparison, Amazon EC2 cloud instances \cite{AmazonEC2docs} and Argonne National Laboratory's Blues HPC (ANLBlues) were deployed to solve the SCUC. Amazon EC2 is a cloud platform that provides rapid scalability of computational resources \cite{AmazonEC2docs}. The local
and cloud infrastructures are summarized in Table I. The Amazon C4 instances are equipped with high-performance processors ideal for computationally intensive applications, whereas the M4 instances provide an overall balance of computing, memory, and network resources. 

\subsection{Computational performance on the cloud}
\begin{figure}[tb]
    \centering
    \includegraphics[width=\linewidth]{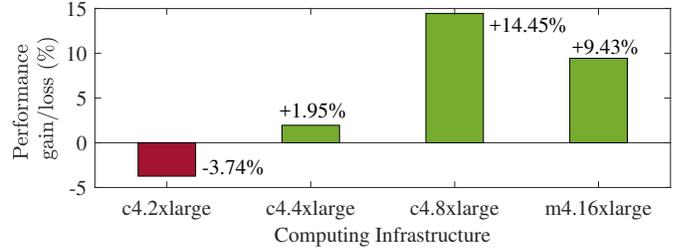}
    \caption{Performance gain or loss (\%). For each instance, the percent change was determined using the average computation time over all trials on Amazon EC2 against the average computation time on ANLBlues. }
    \label{fig:solvetime}
    \vspace{-0pt}
\end{figure}

Since cloud instances are shared resources, the computational resource availability at any given time may be different. Therefore, to obtain an average solve time, Monte Carlo trials were performed, where the UC was solved for 100 trials on both cloud and local ANLBlues instances. To analyze the performance gain or loss, the average percent change was calculated by comparing each cloud instance against the ANLBlues's average. Fig. \ref{fig:solvetime} shows the average percent performance gain or loss for each Amazon EC2 instance.

From Fig. \ref{fig:solvetime}, compared to ANLBlues, c4.2xlarge performs at a computational performance loss due to 8 CPUs compared to the 16 CPUs available within ANLBlues. However, as the number of available CPUs are increased, the C4 family of instances (i.e., c4.4xlarge and c4.8xlarge) provide a positive performance gain compared to ANLBlues. The increased performance between c4.4xlarge to c4.8xlarge indicates that marginal increase in CPUs leads to computational time savings. However, m4.16xlarge, which includes 64 CPUs and 256 GB RAM, is outperformed by c4.8xlarge by +5.02\% with 36 CPUs and 60 GB RAM. The M4 family of instances are not tuned primarily for computation-intensive applications, rather they are for balanced applications requiring compute, high memory, and network resources. 

An analysis such as done in Fig. \ref{fig:solvetime} must be performed by SOs to determine the performance gain unique to the cloud provider of their choice. With the reduction in solve time, the SO may keep the UC problem as-is, change the time intervals by increasing the time horizon (i.e, 24-hour to 36-hour look-ahead) or decreasing the granularity (e.g., 1-hour to 30-minutes), or adding enhancements such as uncertainty management of renewable resources.

\section{Conclusions}

This objective of this work was to introduce the benefits and challenges of outsourcing a highly-complex and crucial application, Unit Commitment (UC), to the cloud. By using the cloud, the SO receives up to a 14.5\% savings in the computation time.  While their are significant benefits to the cloud, the cybersecurity challenges will hinder its widespread adoption, thus mechanisms need to be enacted to secure UC-related data and the infrastructures involved.

\bibliographystyle{IEEEtran}
\bibliography{./cloud.bib}

\end{document}